\documentclass[12pt]{article} 
\usepackage{amsmath}
\usepackage{amssymb}
\usepackage{epsfig}
\usepackage{mcite}

\newcommand{\be}{\begin{equation}}
\newcommand{\ee}{\end{equation}}
\newcommand{\ba}{\begin{eqnarray}}
\newcommand{\ea}{\end{eqnarray}}

\newcommand{\mev    }{\ensuremath{\mathrm{MeV}}}

\newcommand{\gev    }{\ensuremath{\mathrm{GeV}}}
\newcommand{\gevsq  }{\ensuremath{\mathrm{GeV^2}}}
\newcommand{\evsq  }{\ensuremath{\mathrm{eV^2}}}
\newcommand{\km     }{\ensuremath{\mathrm{km}}}
\newcommand{\fb     }{\ensuremath{\mathrm{fb}}}
\newcommand{\der}{\ensuremath{{\operatorname{d}}}}

\newcommand{\numu}{\ensuremath{\nu_\mu}}
\newcommand{\QSav}{\ensuremath{<Q^2>}}
\newcommand{\DMS}{\ensuremath{\Delta M^2}}
\newcommand{\ER}{\ensuremath{E^\prime +\tfrac{\DMS + Q^2}{2 M}}}
\newcommand{\ERav}{\ensuremath{E^\prime +\tfrac{\DMS + \QSav}{2 M}}}
\newcommand{\fSK}{\ensuremath{f_{\rm SK}}}
\newcommand{\ormu}{\ensuremath{1\rm{R}\mu}}
\newcommand{\mmu}{\ensuremath{m_{\mu}}}
\newcommand{\Pmu}{\ensuremath{P_{\mu}}}
\newcommand{\zmu}{\ensuremath{\cos \Theta_{\mu}}}

\begin{document} 
\setlength{\parskip}{0.45cm} 
\setlength{\baselineskip}{0.75cm} 

%
%
%

\begin{titlepage} 
\setlength{\parskip}{0.5cm} 
\setlength{\baselineskip}{0.5cm} 
%
%
\rightline{hep-ph/0307223}
\rightline{DO-TH 2003/12}
\rightline{DESY 03-095}
\vspace*{1cm}

\begin{center} 
\Large
{\bf Muon Spectra of Quasi-Elastic and 1-Pion 
Production Events in LBL Neutrino Oscillation Experiments}
\end{center} 
\vspace{0.5cm}

\begin{center}
{E.\ A.\ Paschos$^1$, D.\ P.\ Roy$^{1,2}$, I.\ Schienbein$^3$, J.-Y.\ Yu$^1$}
\\
\vspace{0.3cm}
{\it $^1$ Theoretische Physik III, 
Univ.\ Dortmund, 44221 Dortmund, Germany}
\\ 
\smallskip
{\it $^2$ Tata Inst.\ of Fundamental Research, 
Bombay-400005, India}
\\
\smallskip
{\it $^3$ DESY/Univ.\ Hamburg, Notkestrasse 85, 
22603 Hamburg, Germany}\\
\end{center}

\medskip
\begin{abstract} 
\noindent 
The muon energy spectra of the quasi-elastic and 1-pion production events 
in a LBL experiment, like K2K,
are predicted to follow closely the neutrino
energy spectrum, with downward shifts of the energy scale
by $\QSav/2 M$ and $(\QSav + M_\Delta^2 - M^2)/2 M$ respectively.
These predictions seem to agree with the observed muon spectra in the
K2K nearby detector.
The corresponding muon spectra in the far-away (SK) detector are
predicted to show characteristic spectral distortions induced by
$\numu$ oscillation. Comparison of the predicted spectral
distortions with the observed muon spectra of the 1-Ring and
2-Ring muon events in the SK detector will help to determine the 
oscillation parameters. The results will be
applicable to other LBL experiments as well.
\end{abstract} 
\end{titlepage} 
 
Recently the KEK to Kamioka long-baseline neutrino
oscillation experiment (K2K) has published its
result \cite{Ahn:2002up}, which confirms the existence
of $\numu$ oscillation as seen in the Super-Kamiokande
(SK) atmospheric neutrino data \cite{Fukuda:1998mi}.
It uses an accelerator-produced $\numu$ beam, which
has a peak energy of $1.2\ \gev$ and a spread of about
$1\ \gev$ on either side -- the full width at half maximum
being $\sim 1\ \gev$ \cite{Ahn:2002up,Itow:NuInt01}.
The beam passes through a nearby neutrino detector (ND),
comprising of a 1 kiloton water Cerenkov (1KT) and a fine-grained
detector (FGD) system, and then travels a distance of $250\ \km$
to enter the SK 50 kiloton water Cerenkov detector.
The charged- and neutral-current event rates measured in the ND are used
to predict the corresponding event rates at the SK detector
in terms of the oscillation parameters.
They report 56 fully contained (FC) events against the expectation
of 80 without oscillation.
The observed deficit agrees well with the neutrino mass and mixing
angles deduced from the atmospheric neutrino 
oscillation data \cite{Fukuda:1998mi}
\be
\sin^2 2 \theta \simeq 1\quad \text{and} \quad 
\Delta m^2 \simeq 3 \times 10^{-3} \evsq \ .
\label{eq:osc}
\ee 
We estimate that the QE, 1-pion production and multi-pion events constitute 
$35\%, \, 50\%$ and $15\%$ at $E_\nu = 1.2\, {\rm GeV}$.
In this region the $\Delta(1232)$ resonance contribution 
accounts for $80-90\%$ of the 1-pion events \cite{Sakuda:NuInt02}, 
with the remainder produced 
from the $P_{11}(1440),\, S_{11}(1535)$ and $D_{13}(1520)$ resonances.
%

In this note we use kinematic considerations to predict
the muon energy spectra of the QE and 1-Pion 
resonance production events which constitute the bulk of the
charged-current $\numu$ scattering events in the K2K experiment.
These predictions can be checked with the observed muon energy
spectra from the ND. We also present the distortion of these muon
spectra due to $\numu$ oscillation, which one expects to see at
the SK detector. Comparison of the predicted muon spectra with 
those of the observed QE and 1-Pion events at the SK detector will be 
very useful in determining the oscillation parameters.

Since the $\Delta$ resonance width of $120\ \mev$ is very small
compared to the spread of the $\nu$ beam energy, one can 
make a narrow resonance approximation for $\Delta$. A common
property of QE and narrow resonance production processes is a
highly restricted kinematics, i.e.,
\be
W^2 = M^2 - Q^2 + 2 M E - 2 M E^\prime \ ,
\label{eq:w2}
\ee
where $W$ represents the nucleon mass $M$ for QE and $M_R$ for
resonance production.
The other variables are as usual the neutrino energy ($E$), the
muon energy ($E^\prime$) and the 4-momentum squared ($Q^2$) transferred
between them.
We can approximate the production cross-section for the narrow
resonance as
\be
\frac{\der \sigma^R}{\der Q^2 \der W^2} = \frac{\der \sigma^R}{\der Q^2}
\ \delta(W^2 - M_R^2)\ ,
\label{eq:xs1}
\ee
where we have replaced the Breit-Wigner factor by the $\delta$-function.
Note that this relation holds exactly for QE scattering with $M_R = M$.
Thus we get from eqs.\ \eqref{eq:w2} and \eqref{eq:xs1}
\be
\frac{\der \sigma^R}{\der Q^2 \der E^\prime} = \frac{\der \sigma^R}{\der Q^2}
\ \delta(E^\prime - E + \tfrac{\DMS + Q^2}{2 M})\ ,
\label{eq:xs2}
\ee
where $\DMS = M_R^2 - M^2$, which is 0 for QE scattering.
Convoluting this cross-section with the neutrino beam spectrum
$f(E)$ gives the flux averaged cross-section
\be
\int f(E)\ \frac{\der \sigma^R}{\der Q^2 \der E^\prime}\ \der E 
= \frac{\der \sigma^R}{\der Q^2}(E=E_R)\ f(E=E_R)\ , \
E_R = \ER \ .
\label{eq:xsf}
\ee
Moreover to a first approximation one can integrate over $Q^2$ to
write
\be
\int f(E)\ \frac{\der \sigma^R}{\der E^\prime}\ \der E 
= \sigma^R(\ERav)\ f(\ERav)\ , 
\label{eq:xsfav}
\ee
where $\QSav \simeq 0.3\ \gevsq$ for both QE and $\Delta$ production
and the neutrino energy $E = 0 - 3.5\ \gev$ for the K2K experiment
\cite{Ahn:2002up}.
It assumes the effective range of $Q^2$ integration to be small, so that
over the corresponding neutrino energy range 
$\Delta E = \Delta Q^2/2 M$ the $f(E)$ and $\sigma^R(E)$ can be 
approximated by their mean values.
Thus we see from eq.\ \eqref{eq:xsf} or \eqref{eq:xsfav} that for 
both QE and narrow resonance production the shape of the muon
spectrum is predicted to follow the shape of the incident 
neutrino spectrum, with a downward shift of the energy scale by
$\tfrac{\DMS + Q^2}{2 M}$.
One notes an interesting analogy between this result 
and the quark parton model, 
where the structure function in deep inelastic scattering 
follows the shape of the parton energy
distribution within the nucleon.

Of course the shape of the muon spectrum is not exactly identical
to the shape of the neutrino spectrum, because it is modulated by
the QE (Resonance production) cross-section. The QE ($\Delta$ production) 
cross-section shows a threshold rise over $E = 0 - 1\ \gev$
($0.4 - 1.4\ \gev$) and becomes flat thereafter.
This will be reflected in steepening of the muon spectra at the
low energy end, as we shall see below. On the other hand the shape
of the muon spectra on the higher energy side of the peak
will be practically identical to that of the incident
neutrino spectrum for both QE and $\Delta$ production processes.
Therefore the relative size of the SK to the ND cross-sections over this
region provides a direct measure of the spectral distortion and hence the
underlying oscillation parameters.

These are simple but very useful results, particularly since
$\Delta$ accounts for about $80 - 90 \%$ of the 1-Pion production
cross-section and the 
QE plus 1-Pion production together account for $85 \%$ of the
charged-current $\numu$ cross-section at the K2K experiment.
It means that one can predict the muon energy spectra of the
QE and 1-Pion events in terms of the neutrino energy spectrum
from general kinematic considerations without practically any
model dependent ansatz. These predictions can be verified with
the muon spectra of the QE and 1-Pion events observed by their ND.
Even more importantly one can then predict the distortions in the
corresponding muon energy spectra at the SK detector, induced
by $\numu$ oscillation. Thus the predictions corresponding to
eqs.\ \eqref{eq:xsf} and \eqref{eq:xsfav} for the SK detector are
\be
\int \fSK(E)\ \frac{\der \sigma^R}{\der Q^2 \der E^\prime}\ \der E 
= \frac{\der \sigma^R}{\der Q^2}(E_R)\ f(E_R)\ 
[1- \sin^2 2 \theta\ \sin^2 (\tfrac{1.27 \Delta m^2 L}{E_R})]\ ,
\label{eq:SK}
\ee
with $E_R = \ER$, and
\be
\int \fSK(E)\ \frac{\der \sigma^R}{\der E^\prime}\ \der E 
= \sigma^R(E_R)\ f(E_R)\ 
[1- \sin^2 2 \theta\ \sin^2 (\tfrac{1.27 \Delta m^2 L}{E_R})]\ ,
\label{eq:SKav}
\ee
with $E_R = \ERav$.
Comparing these predictions with the observed muon energy spectra
of the QE and 1-Pion events of the SK detector will test the 
spectral distortion due to $\numu$ oscillation and determine the
oscillation parameters.

\begin{figure}[htb]
\centering
\includegraphics[angle=0,width=10cm]{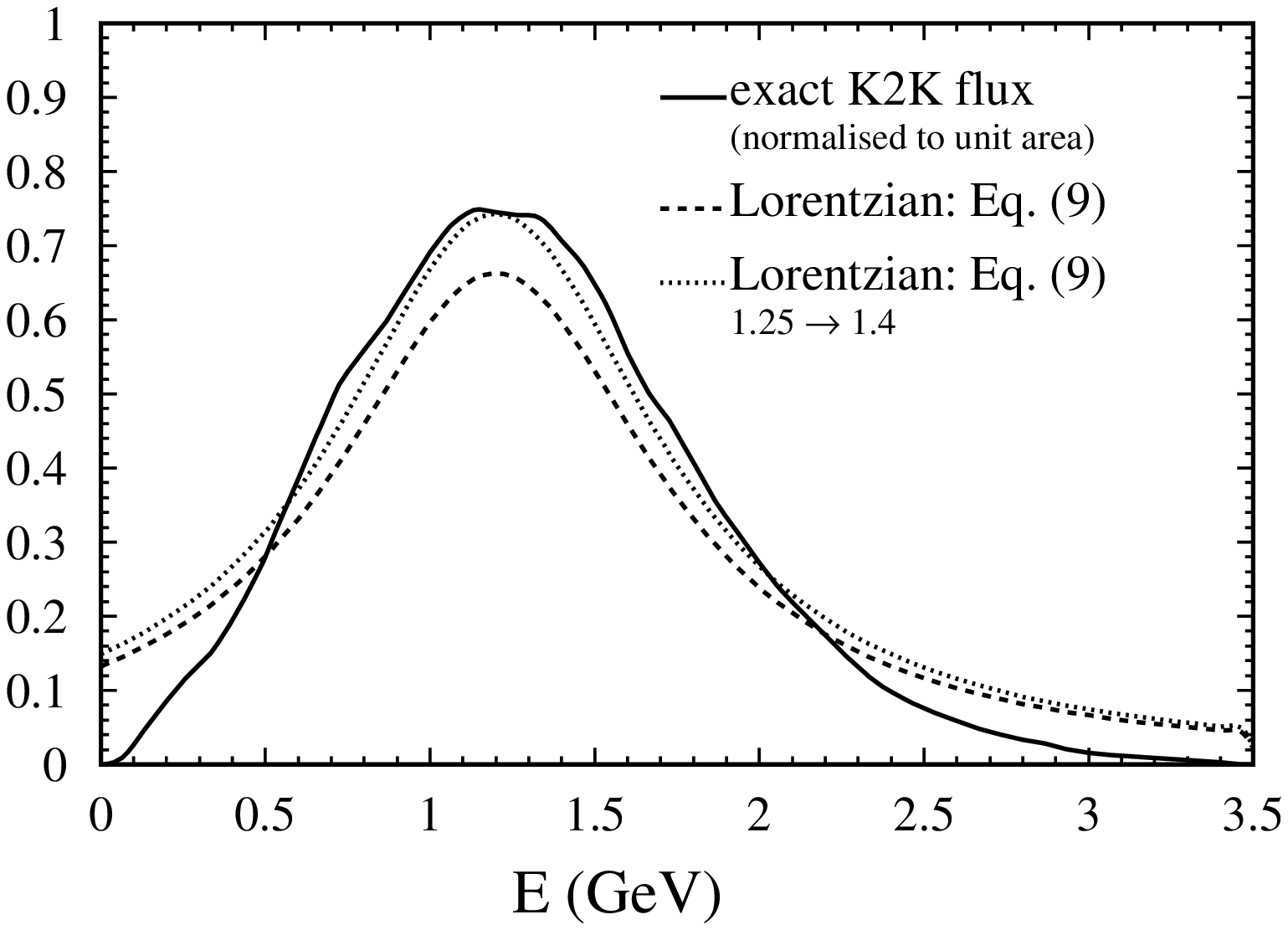}
\vspace*{-1.5cm}
\caption{\sf K2K neutrino energy spectrum. The solid line is the exact
spectrum normalised to unit area. The dashed line shows the 
approximated spectrum by the Lorentzian in Eq.~\eqref{eq:flux}. 
We also show for comparison as dotted line the same Lorentzian, 
but with normalisation 1.4 instead of 1.25.}
\label{fig:k2k_spectrum}
\end{figure}

Let us first present a simple analytic form of QE and $\Delta$ 
production cross-sections for the ND and SK detector corresponding
to eqs.\ \eqref{eq:xsfav} and \eqref{eq:SKav} respectively.
For this purpose we have approximated the neutrino energy spectrum
by the Lorentzian
\be
f(E) = \frac{1.25}{\pi}\frac{\gamma}{(E-E_0)^2 + \gamma^2}\ , \ 
E_0 = 1.2\ \gev\ , \ \gamma = 0.6\ \gev\ ,
\label{eq:flux}
\ee
which agrees with the shape of the K2K neutrino energy spectrum quite
well. The normalization factor of 1.25 ensures that the total flux is
normalized to 1 when integrated over the experimental range of 
$E = 0 - 3.5 \ \gev$ instead of $-\infty$ to $\infty$.
Figure \ref{fig:k2k_spectrum} shows a comparison of the K2K neutrino
energy spectrum with the parametrisation of eq.~\eqref{eq:flux}.
Substituting this in eqs.\ \eqref{eq:xsfav} and \eqref{eq:SKav}
gives
\be
\int f(E)\ \frac{\der \sigma^R}{\der E^\prime}\ \der E 
= \sigma^R(\ERav)\ 
\frac{1.25}{\pi}\frac{\gamma}{(\ERav-E_0)^2 + \gamma^2}
\ , 
\label{eq:lorentz1}
\ee
and
\be
\int \fSK(E)\ \frac{\der \sigma^R}{\der E^\prime}\ \der E 
= \sigma^R(\ERav)\ 
\frac{1.25}{\pi}\frac{\gamma \left[1- \sin^2 2 \theta\ 
\sin^2 \left(\tfrac{1.27 \Delta m^2 L}{\ERav}\right)\right]}
{(\ERav-E_0)^2 + \gamma^2}\ .
\label{eq:lorentz2}
\ee
Finally we have approximated the QE and $\Delta$ production
cross sections by
\ba
\sigma^N(E) &=& E_N \times 4.5\ \fb \ , \ E_N = \min(E,1)\ \gev\ ,
\label{eq:QE}
\\
\sigma^\Delta(E) &=& (E_D - 0.4\ \gev) \times 5.5\ \fb \ , 
\ E_D = \min(E,1.4)\ \gev\ .
\label{eq:RES}
\ea
They correspond to linear threshold rise for the QE ($\Delta$ production)
cross-section between 0--1 (0.4--1.4) $\gev$, which describes the
cross-sections quite well. The normalizations correspond to the
average cross-section per nucleon for a $H_2O$ target \cite{Paschos:2000be}.

For the sake of completeness we have also calculated the $P_{11}(1440)$ and
$S_{11}(1535)$ contributions using the narrow resonance
approximation. Of course it may not work so well for the $P_{11}(1440)$,
which has a width of 350 MeV. But since these resonance contributions to
the 1-Pion production cross-section are small, the resulting error will
be insignificant. Although there are significant differences between the
parametrisations of ref.\ \cite{Paschos:2000be,*Paschos:2001np,Paschos:2002mb} 
and ref.\ \cite{Rein:1981wg} for the individual
resonances, the cumulative contributions are very similar for the two
sets of parameters. We have used the parametrisation of 
ref.\ \cite{Paschos:2000be,*Paschos:2001np,Paschos:2002mb}
for these resonance production cross-sections.

\begin{figure}[htb]
\centering
\vspace*{-2.5cm}
\includegraphics[angle=0,width=7.9cm]{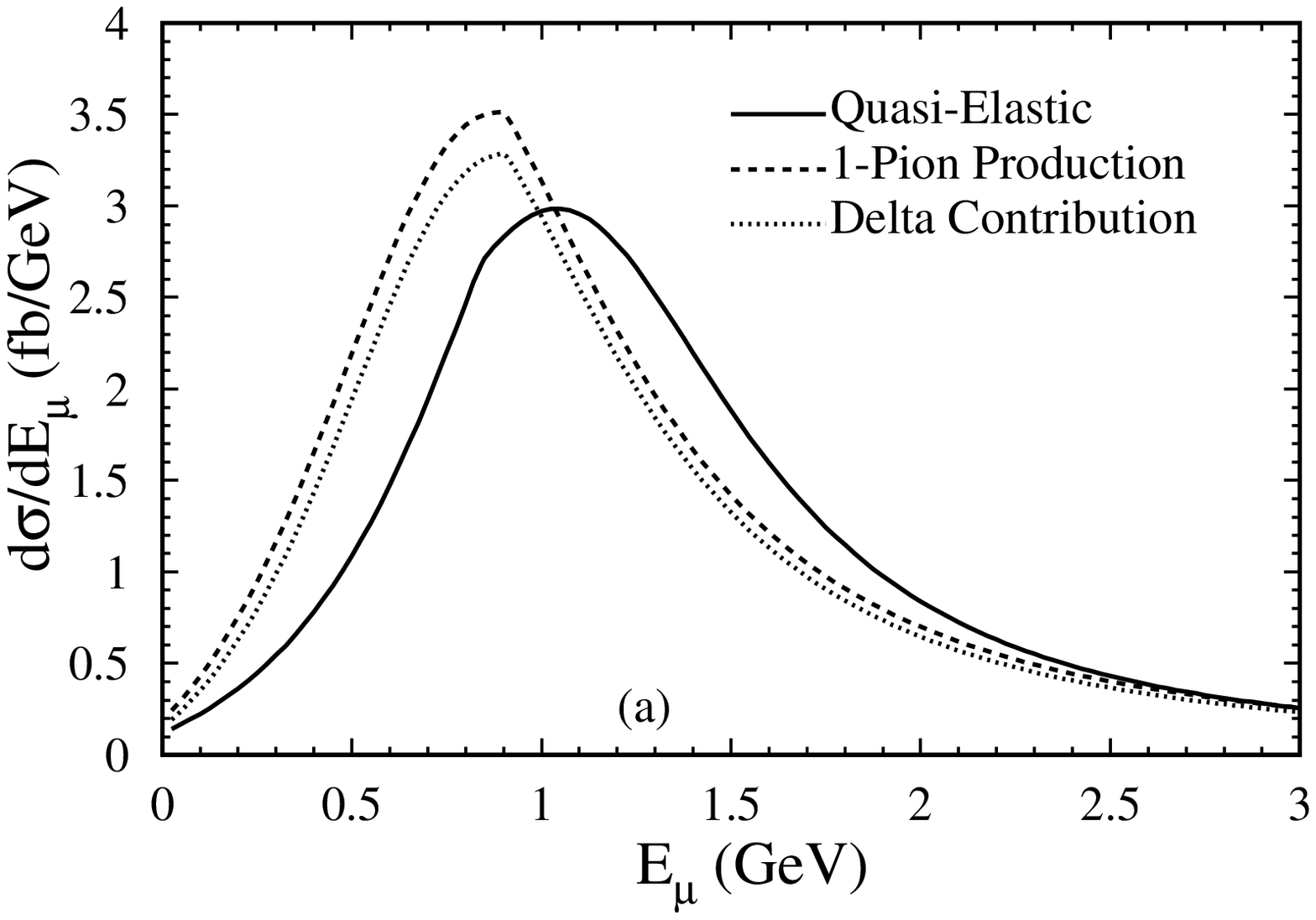}
\includegraphics[angle=0,width=7.9cm]{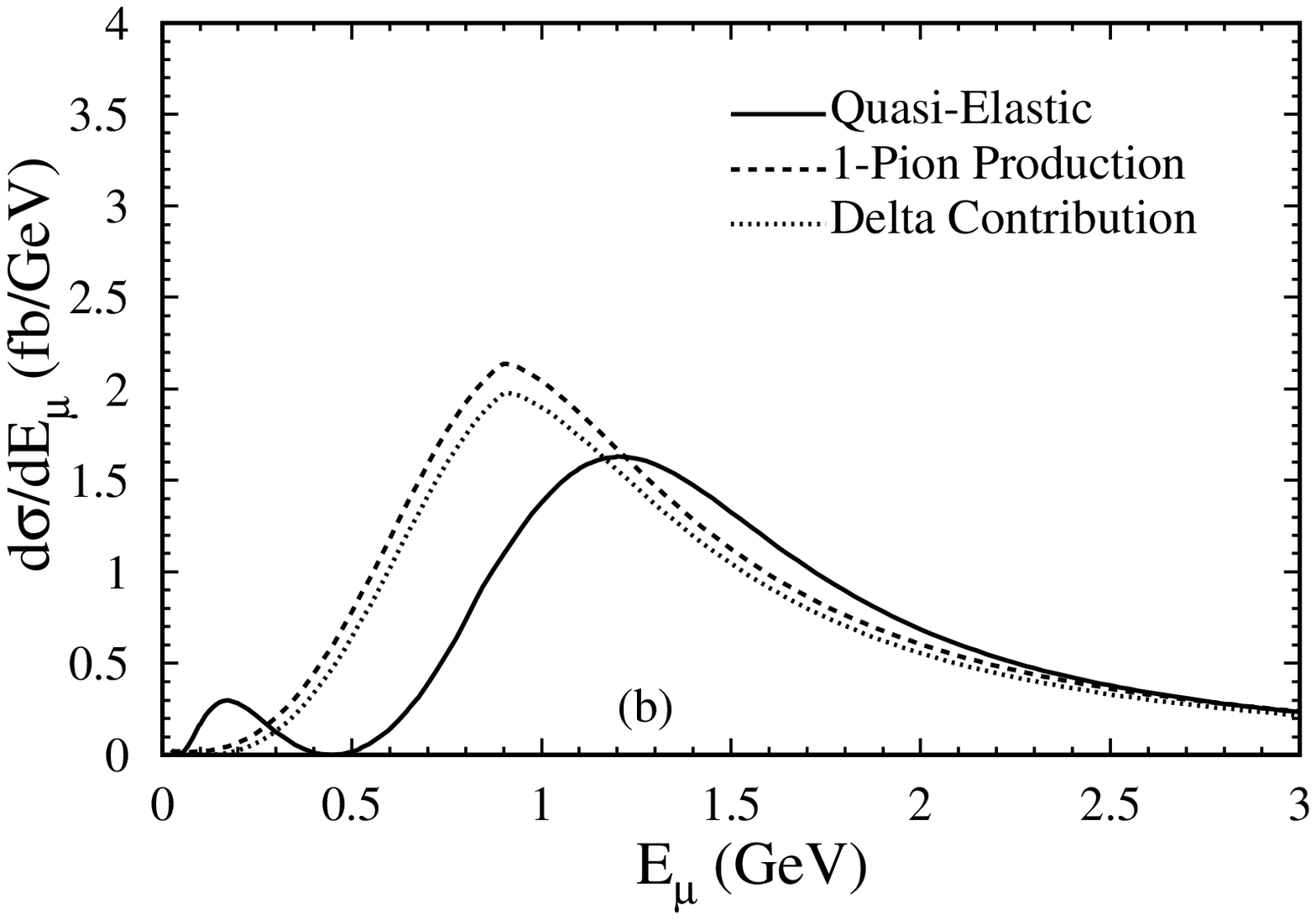}
\vspace*{-1.5cm}
\caption{\sf Approximate prediction of the muon energy spectra for the (a) Nearby and (b) SK detectors of the K2K experiment. The Quasi-Elastic and the 1-pion production cross-sections are shown along with the $\Delta$ contribution to the latter.}
\label{fig:2}
\end{figure}

Figure \ref{fig:2}a shows the predicted muon energy
spectra for the QE and 1-Pion events, the latter being dominated by
the $\Delta$ contribution (shown separately).
One can clearly see the downward shift of the peak from $1.2\ \gev$
by $\QSav/2 M \simeq 0.15\ \gev$ for the QE and 
$(\DMS+\QSav)/2 M \simeq 0.4 - 0.5\ \gev$ for the $\Delta$ contribution.
One can also see the reflection of the threshold rise
from eqs.\ \eqref{eq:QE} and \eqref{eq:RES}, particularly for the
$\Delta$ contribution. On the other hand the muon energy spectra
to the right of the peak closely follow the shape of the neutrino
energy spectrum of 
Fig.\ \ref{fig:k2k_spectrum}.
Both the curves agree reasonably well with the more exact calculation of QE and
1-Pion production cross-sections, except for a small range near $E^\prime\simeq 0.4\ \gev$
as discussed  below.
One can see them to be in general 
agreement with the observed muon spectra of QE and non-QE events
of the K2K ND data \cite{Ahn:2002up}. 
But a quantitative comparison will need
to include things like the detection efficiency and experimental
cuts, which is beyond the scope of the present work.

Figure \ref{fig:2}b
shows the corresponding muon energy spectra of the QE and 1-Pion
events for the 
far away
detector, predicted by 
eqs.\ (\ref{eq:lorentz2}, \ref{eq:QE}, \ref{eq:RES}) along with
a small contribution from the higher resonances.
One can clearly see the distortion of the muon energy spectra
due to $\numu$ oscillation, where we have used the oscillation
parameters of eq.\ \eqref{eq:osc}.
They are again in reasonable agreement with the more exact results
presented below.
It will evidently be very useful to compare these predictions with
the observed muon energy spectra of QE and 1-Pion events from the
SK detector which is not available to us. Again a quantitative
comparison will require taking into account the detection efficiency
of the muon and the pion, which can best be done by the K2K
Collaboration.

We shall now present the results of 
an exact calculation.  
It is based on 
numerical integration
of eqs.\ \eqref{eq:xsf} and \eqref{eq:SK}
for QE scattering, while for the resonance production it uses the
corresponding formulae including the resonance widths. Apart from the
leading contribution from the $\Delta(1232)$ resonance, it includes
subleading contributions from the $P_{11}(1440)$ and $S_{11}(1535)$
resonances. We estimate the contribution from the
still higher resonances along with the nonresonant background to be no
more than $5 - 10 \%$ of the 1-Pion production cross-section at this energy.
Therefore the accuracy of our prediction 
should be as good as that of the K2K experiment.
The parametrisation of the QE and the resonance production cross-sections
used in this analysis can be found in \cite{Paschos:2000be,*Paschos:2001np,Paschos:2002mb}.

For the excitation of the $\Delta$-resonance we use the 
formalism of \cite{Adler:1968tw,Schreiner:1973mj} with two 
distinct form factors for the vector and axial currents.
The vector and the axial vector form factors are 
modified dipoles \cite{eap:2003}.
For the $P_{11}$ and $S_{11}$ resonances we use the 
form factors from \cite{Fogli:1979cz}.

\begin{figure}[htb]
\centering
\includegraphics[angle=0,width=7.9cm]{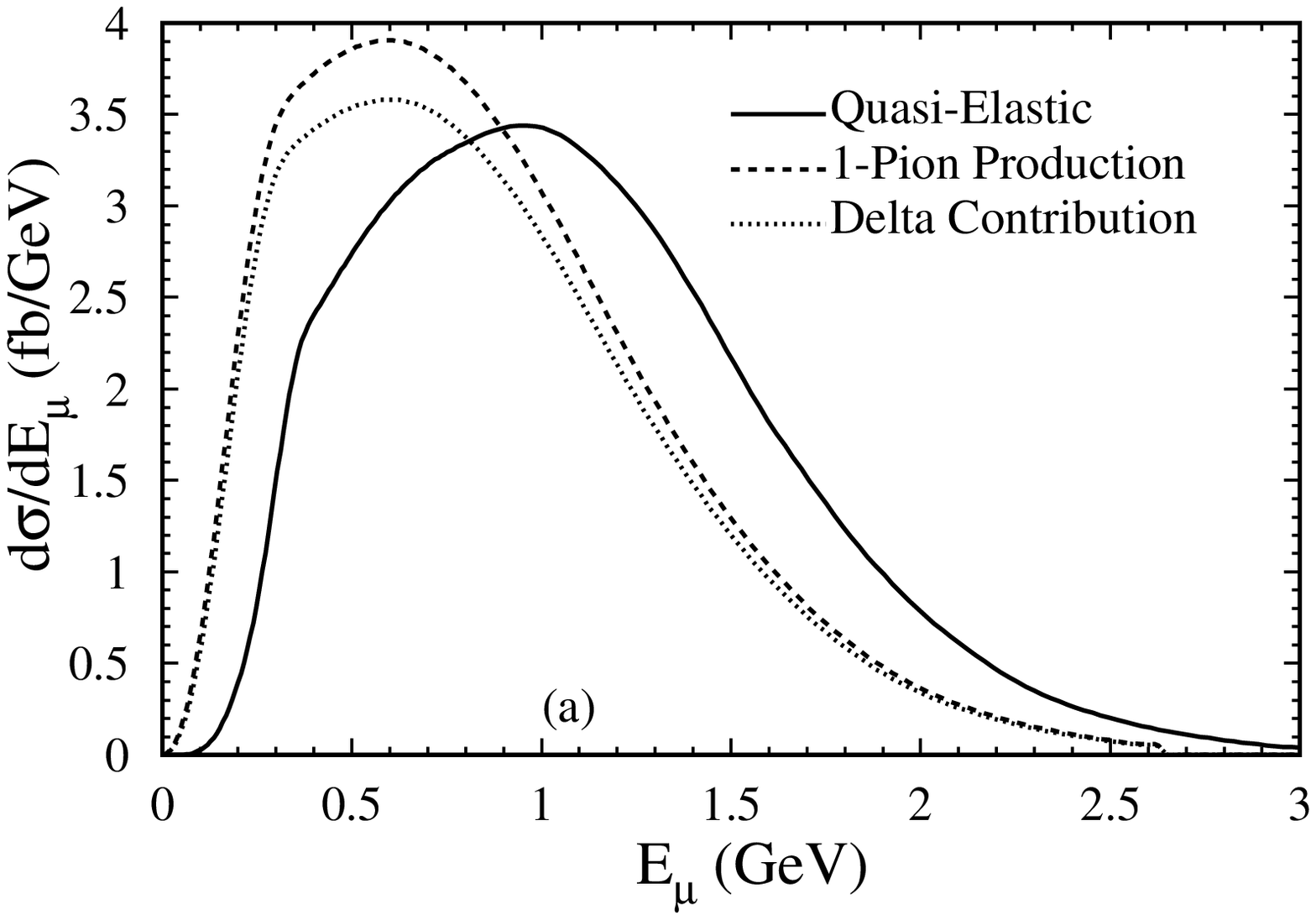}
\includegraphics[angle=0,width=7.9cm]{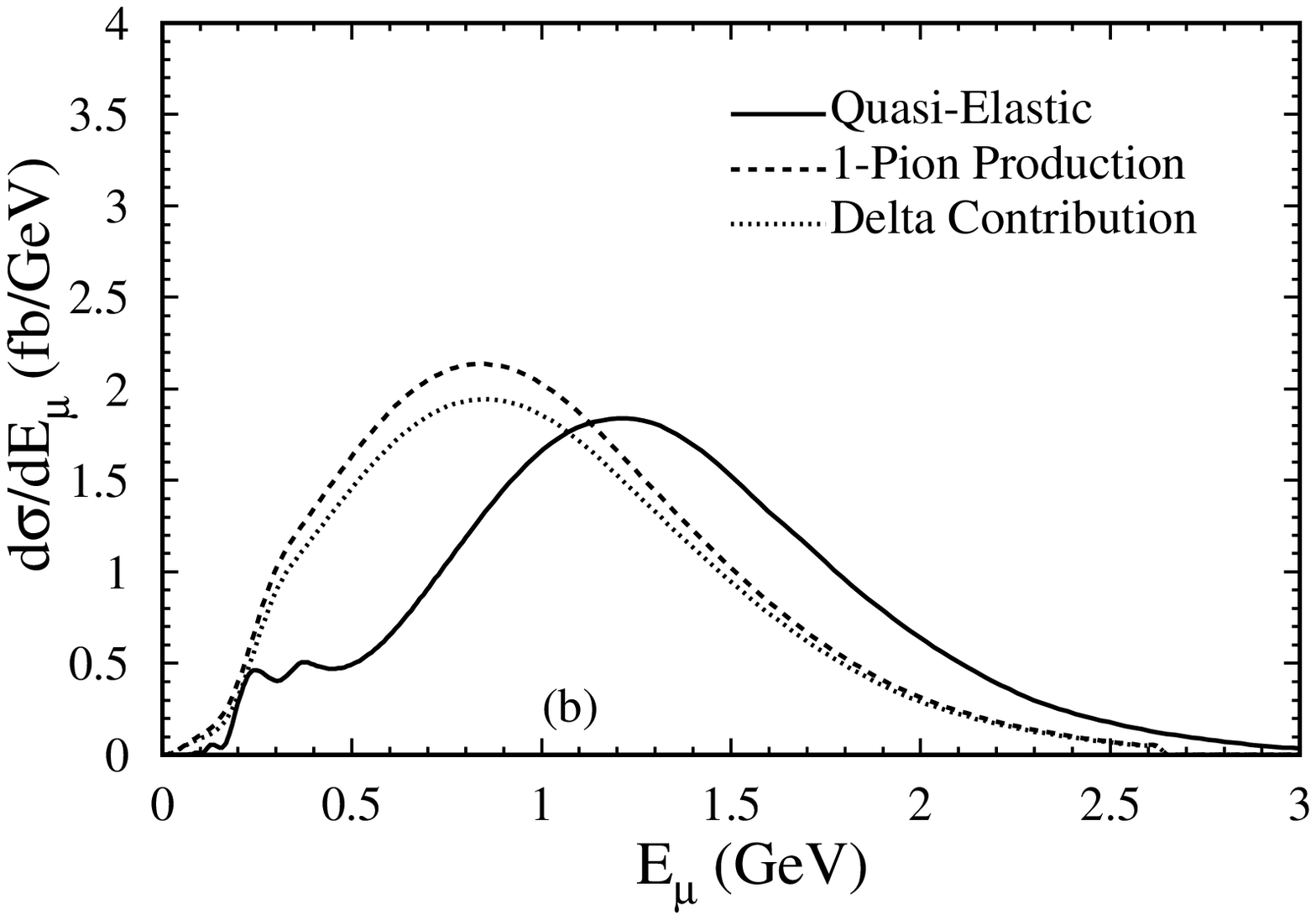}
\vspace*{-1.5cm}
\caption{\sf Exact predictions
 of the muon energy spectra for the (a) Nearby and (b) SK detectors of the K2K experiment. The Quasi-Elastic and the 1-Pion production  cross-sections are shown along with the $\Delta$ contribution to the latter.}
\label{fig:3}
\end{figure}

Fig.\ \ref{fig:3}a  
shows the predicted muon spectra for the QE and 1-Pion
production processes.
One can clearly see the downward shift of the peak from $1.2\ \gev$ by
$\QSav/2 M$ 
for the QE and
$(\DMS+\QSav)/2 M$ for the $\Delta$ contribution, dominating
the 1-Pion production process. The steepness of the muon spectra at the
low energy end reflects the threshold rise. On the other hand they closely
follow the shape of the neutrino energy spectrum on the right side of the peak.
All these features are in general agreement with the approximate results
shown in Fig.\ \ref{fig:2}. One sees however a clear broadening of the
muon spectra below $1\ \gev$ compared to the latter. In particular the exact
calculation for QE shows a clear excess over the approximate one near 
$E^\prime \simeq 0.4\ \gev$ as mentioned earlier. This is because the 
assumption of a small range of $Q^2$ integration made in deriving eq.\ (6)
from eq.\ (5) does not hold here, as the rapid increase of the flux factor 
upto $E = E^\prime + Q^2/2 M = 1.2\ \gev$ stretches the effective range of
$Q^2$ upto $1.5\ \gevsq$. Note that below $E^\prime = 0.4\ \gev$ the
$Q^2$ range is restricted by the kinematic constraint 
$Q^2 \le 4 M E^{\prime 2} / (M - 2 E^\prime)$.
So the rapid fall of the spectrum with $E^\prime$ simply reflects the fall
of the flux and the threshold factors with $E$.

The predicted muon spectra of Fig.\ \ref{fig:3}a agree
reasonably well with the
corresponding spectra of the K2K ND\ \cite{Ahn:2002up} for both the 
QE and the non-QE
parts. In particular one can compare the predicted QE spectrum with their 
simulated QE spectrum shown in Fig.\ 1 of ref.\ \cite{Ahn:2002up}. 
Their Fig.\ 1a and c
show separately the QE muon spectra for the 1-Ring muon ($\ormu$) sample of the
1KT and the QE enhanced sample of the FGD respectively. 
The two play complementary
roles in covering the complete muon energy range, as the 1KT and the FGD have
high efficiencies at $E^\prime < 1\ \gev$ and $E^\prime \gtrsim 1\ \gev$, 
respectively \cite{Ahn:2002up}. 
One can not compare our predicted muon spectra with these figures
quantitatively without folding in these efficiency factors, which
are not available to us.
But
there is good qualitative agreement between the predicted QE spectrum of our
Fig.\ \ref{fig:3}a with their Fig.\ 1c at $E^\prime \gtrsim 1\ \gev$ 
and Fig.\ 1a at
$E^\prime < 1\ \gev$. While the former shows the position of the peak and 
the shape of the spectrum to the right, the latter shows the broadening of
spectrum down to $E^\prime \simeq 0.4\ \gev$ as discussed above. 
Similarly one
sees good agreement between the predicted muon spectrum of our 
Fig.\ \ref{fig:3}a for
1-Pion events with the non-QE spectra of their Fig.\ 1c,d at 
$E^\prime \gtrsim 1\ \gev$
and Fig.\ 1a at $E^\prime < 1\ \gev$. 
Thus one has a simple and robust prediction
for the shape of the muon spectrum in terms of the neutrino spectrum not only
for the QE events but also for the 1-Pion events, which dominate the inelastic events.

Fig.\ \ref{fig:3}b shows the corresponding muon energy spectra of the
QE and 1-Pion events for the SK detector, predicted by eqs.\ \eqref{eq:osc}
and \eqref{eq:SK}.
Again the curves are in reasonable agreement with the approximate results
of Fig.\ \ref{fig:2}b.
One can clearly see the distortion of the muon energy spectrum due the
$\numu$ oscillation.
They should be compared with the observed muon energy spectra of the
1-Ring and 2-Ring muon events at the SK detector, after taking into account
the pion detection efficiency. We hope such a comparison will be done by the
K2K collaboration.

To estimate the pion detection efficiency one 
needs to take into account the
nuclear absorption, Pauli blocking 
and charge exchange taking place during 
the rescattering of the produced pions. 
We have included these effects following the prescription of 
ref.\ \cite{Adler:1974qu}.
Since the dominant contribution to 1-Pion production process comes from
resonance production on Oxygen, we have evaluated the effects of nuclear
absorption and rescattering  on the produced pion for this case.
The relevant subprocesses are
$\nu p \to \mu^- p \pi^+$, $\nu n \to \mu^- n \pi^+$ and 
$\nu n \to \mu^- p \pi^0$ with relative cross-sections $9:1:2$ 
for the dominant contribution from $\Delta$.

\begin{figure}[htb]
\centering
\includegraphics[angle=0,width=7.9cm]{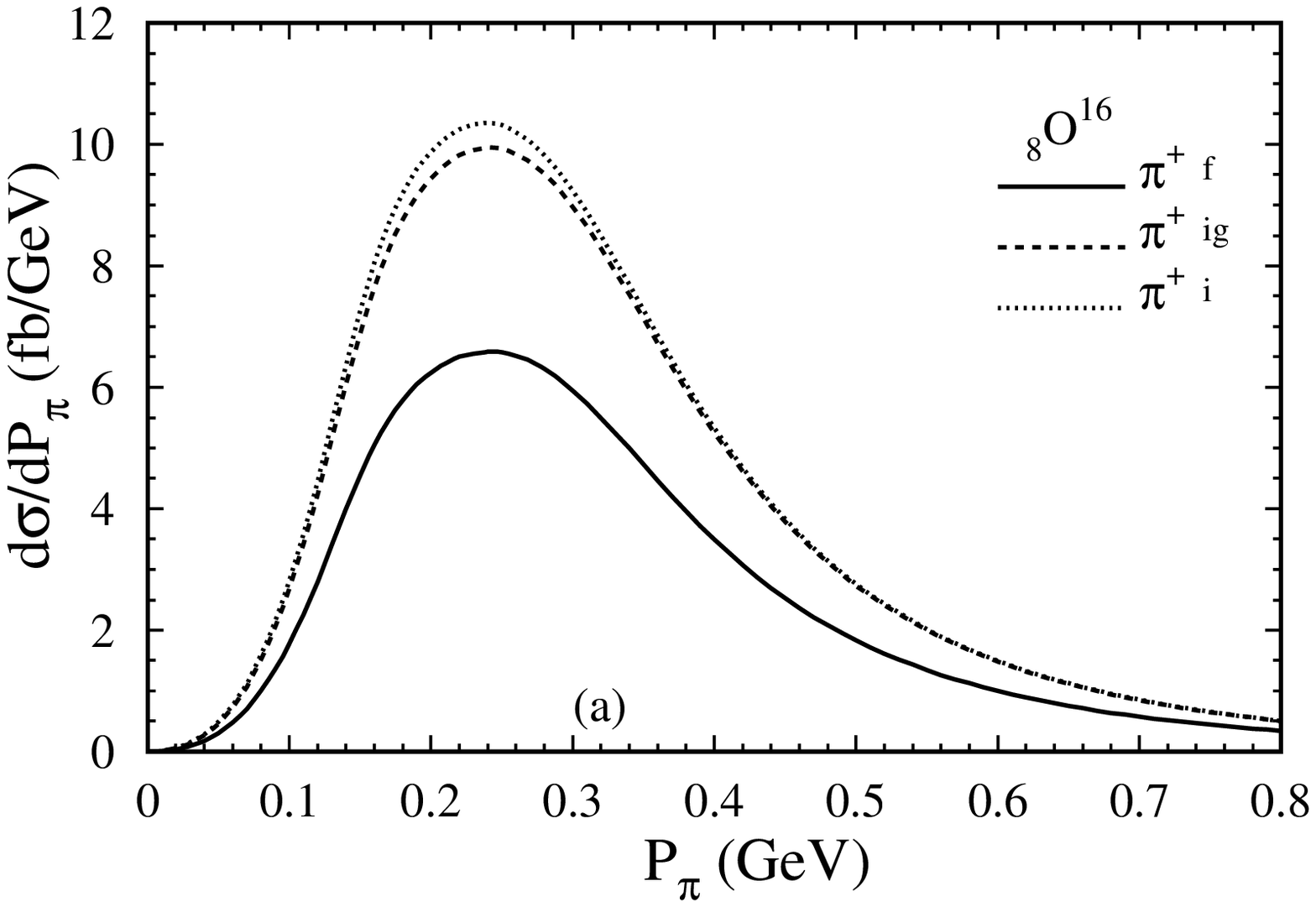}
\includegraphics[angle=0,width=7.9cm]{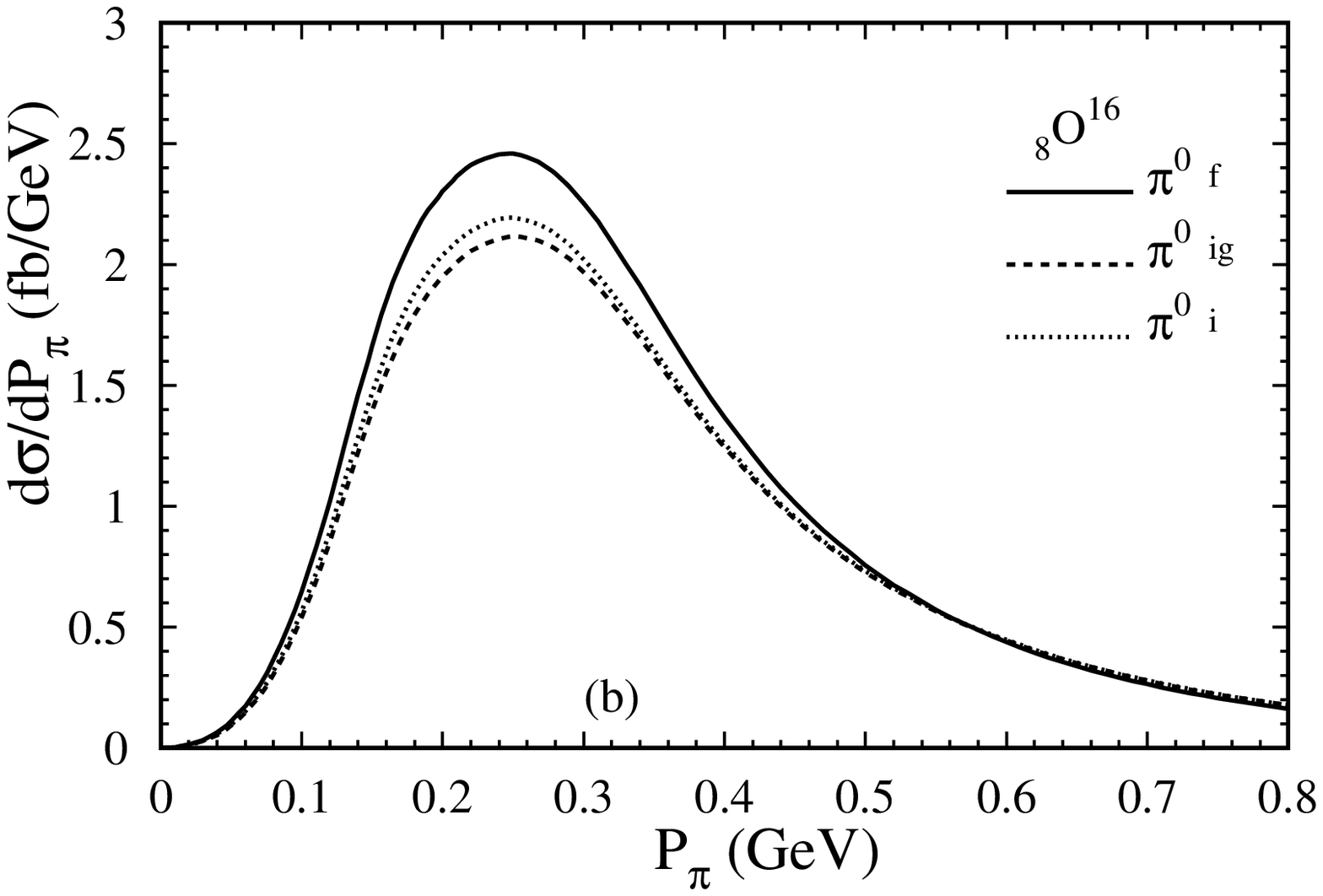}
\vspace*{-1.5cm}
\caption{\sf The momentum distribution of the decay pion for resonance production by charged current in Oxygen before (dotted) and after nuclear correction (solid). The dashed line shows only the effect of Pauli blocking.}
\label{fig:4}
\end{figure}

Fig.\ \ref{fig:4} shows the effects of nuclear corrections on the 
produced $\pi^+$ and $\pi^0$ spectra from these processes 
for the nearby detector averaged over the neutrino spectrum. 
The results are very similar for the SK detector.
Nuclear rescattering effects result in enhancing the $\pi^0$ events
at the cost of the dominant $\pi^+$ component. But taken together we see
a nearly $20 \%$ drop in the rate of 1-Pion events due to nuclear absorption
of the produced pion.
Moreover about $10 \%$ of the remaining events corresponds 
to the pion momentum
being less than the Cerenkov threshold of $100\ \mev$. Therefore one
expects about $70\%$ of the $\Delta$ events to give a detectable
pion ring at the SK detector while the remaining $30\%$ appears like
a QE event.
Adding the latter to the $35 \%$ of genuine QE events would imply
that about $50 \%$ of the CC events will appear QE-like at the SK detector.
This seems to agree with the observation of $29$ 1-Ring muon
($\ormu$) events at the SK detector out of their $56$ FC events.
It also agrees with their Monte Carlo simulation of $44$ $\ormu$
events out of the total of $80$ FC events in case of no oscillation.
Thus the observed muon energy spectrum of the $29$ $\ormu$ events
should be compared with a weighted average of the predicted spectra
for QE and 1-Pion events. Alternatively the observed muon energy spectrum
of the sum of 1-Ring and 2-Ring muon events could be compared with
the predicted spectrum of the sum of QE and 1-Pion events.
Although a part of the 2-Ring events may come from multi-pion
production, the resulting error may be small since multi-pion events at 
the ND constitute only $\sim 15 \%$ of CC events.

It should be mentioned here that the K2K Collaboration has reconstructed the
neutrino energy for the $29$ $\ormu$ events using the kinematic relation
\be
E = \frac{M E^\prime - \mmu^2/2}{M - E^\prime + \Pmu \zmu}
\label{eq:E-reconstruction}
\ee
and used it for testing the spectral distortion due to neutrino oscillation
\cite{Ahn:2002up}.
While this relation holds for genuine QE events, it is not possible
to reconstruct the neutrino energy for an inelastic event using only
the muon parameters.
As we have seen above, about $1/3$rd of the $29$ $\ormu$ events
may come from 1-Pion production. Hence we can see no reliable
way of reconstructing the neutrino energy for this $\ormu$ sample on an
event by event basis.
On the other hand the muon energy is a directly measurable quantity
for each event. Therefore it seems to us to be a better variable for
testing the spectral distortion phenomenon compared to the reconstructed
neutrino energy.

Let us note finally that the above formalism is applicable not only
to K2K but also to future long baseline experiments
like J2K, MINOS and the CERN-Gran Sasso experiments which plan to
use low energy $\numu$ beams \cite{Itow:NuInt01,Lipari:NuInt01}.
Therefore it will be very useful to extend this analysis for the
beam energy spectra and the target nuclei of these
experiments.

\vspace{2.0cm}
\noindent{\large\bf{Acknowledgement}}

The support of the 
``Bundesministerium f\"ur Bildung, Wissenschaft, Forschung und
Technologie'', Bonn under contract 05HT1PEA9 is gratefully 
acknowledged.
The work of D. P. Roy was supported by BMBF.
We wish to thank Dr.\ M.\ Sakuda for helpful correspondence.


\end{document}